\documentclass[runningheads]{llncs}
\usepackage{graphicx}

\usepackage{amsmath}
\usepackage{amsfonts}
\usepackage{hyperref}
\usepackage{makecell}
\usepackage{url}
\usepackage{float}
\usepackage{kantlipsum}
\allowdisplaybreaks
\begin{document}
\setlength{\belowcaptionskip}{-4pt}
\title{Hierarchical Uncertainty Estimation for\\ Medical Image Segmentation Networks}
\titlerunning{VAE U-Net}
%
\author{Xinyu Bai\thanks{\email{mb220@ic.ac.uk}, Work done as Master's project at Department of Computing, Imperial College London} \and Wenjia Bai\thanks{\email{w.bai@ic.ac.uk}, Department of Computing and Department of Brain Sciences, Imperial College London}}
\authorrunning{}

\institute{}
\maketitle

\begin{abstract}
Learning a medical image segmentation model is an inherently ambiguous task, as uncertainties exist in both images (noise) and manual annotations (human errors and bias) used for model training. To build a trustworthy image segmentation model, it is important to not just evaluate its performance but also estimate the uncertainty of the model prediction. Most state-of-the-art image segmentation networks adopt a hierarchical encoder architecture, extracting image features at multiple resolution levels from fine to coarse. In this work, we leverage this hierarchical image representation and propose a simple yet effective method for estimating uncertainties at multiple levels. The multi-level uncertainties are modelled via the skip-connection module and then sampled to generate an uncertainty map for the predicted image segmentation. We demonstrate that a deep learning segmentation network such as U-net, when implemented with such hierarchical uncertainty estimation module, can achieve a high segmentation performance, while at the same time provide meaningful uncertainty maps that can be used for out-of-distribution detection. The code is publicly available \href{https://github.com/BXYMartin/Python-Uncertainty\_Aware\_Vision\_Transformer}{here}.

\keywords{Medical image segmentation \and hierarchical image representation \and uncertainty modelling \and variational inference.}
\end{abstract}
\section{Introduction}
Image segmentation plays an essential role in medical image analysis, which enables extraction of clinically relevant information about anatomical structures, detection of abnormalities and visualisation of regions of interest. In recent years, automated medical image segmentation has made solid progress with deep learning. Given clinicians' concerns over exclusive reliance on automated segmentations for critical decisions \cite{uqmedicine}, one of the solutions is to corporate uncertainty estimation into the network. This not only improves model interpretability for ambiguous areas, but also serves as a safeguard for unexpected inputs.

In this work, we propose a generic hierarchical uncertainty estimation framework, named as VAE U-net, for medical image segmentation task. VAE U-net integrates uncertainty modelling with the skip-connection module between encoder and decoder. Different from previous works \cite{PhiSeg2019,kohl2019hierarchical}, the sampled latent features are directly fed into the decoder without upsampling and then concatenated with features from the skip-connection layer. This may allow the gradients to flow easily to the initial layers of the encoder to learn good representations. We demonstrate that VAE U-net achieves state-of-the-art segmentation performance while captures a very good estimation of the uncertainty as a light-weight model. 

\section{Related Works}
Capturing uncertainty within the learning process is often implemented by model ensemble or by using variational Bayesian methods \cite{Abdar_2021}. For model ensembling, Lakshminarayanan et al. \cite{lakshminarayanan2017simple} introduced a randomisation-based approach to ensemble multiple networks and showed that majority voting using these networks greatly improves the accuracy. A more advanced technique to produce diversified weight samples was proposed in \cite{zhao2022efficient}, which includes cyclical learning rate and multi-modal checkpoint ensemble. Limitations for model ensemble include reduced interpretability as well as additional costs in computation time. 

Apart from model ensemble, variational inference provides another way for estimating uncertainty \cite{graves2011practical}. Dropout \cite{srivastava2014dropout}, initially proposed as a regularization layer, is recognised as a way for approximate inference in a complex network \cite{dropoutasbnn}. Similar techniques were adopted by flipout \cite{wen2018flipout}, which perturbs the weight distribution in mini-batches. More recent works are dedicated to introducing auxiliary modules to an existing network. To directly learn the uncertainty from samples, Kohl et al. \cite{kohl2018probabilistic} included a side branch to the U-net architecture to learn the latent distribution with a conditional variational auto-encoder. Later, architectures that used  hierarchical probabilistic U-net to model ambiguities at multiple scales were proposed \cite{PhiSeg2019,kohl2019hierarchical}, where the uncertainty estimator was interleaved with feature extraction by the decoder.

\section{Methods}
\subsection{Problem Formulation}
\label{section:formulation}
Let $X$ denote the input image and $Y$ denote the segmentation map. The image $X$ is mapped to a latent space $z$ via an encoder $q_\phi(\mathbf{z} \mid X)$, which describes the posterior distribution of $z$ conditioned on $X$. In a variational auto-encoder (VAE), the posterior $q_\phi(\mathbf{z} \mid X)$ is typically constrained by a prior $p(z)$ which is a Gaussian distribution. We assume that the segmentation $Y$ can be mapped to the same latent space $z$ via an encoder $p_\theta(\mathbf{z} \mid Y)$, i.e. the image and the segmentation share an aligned latent space. Based on this assumption, we formulate the following learning objective,
{
\small
\begin{align}
\max_{\theta, \phi} \mathbb{E}_{X, Y\sim D}[\mathbb{E}_{z \sim q_\phi(\mathbf{z} \mid X)}[\log p_\theta(Y \mid \mathbf{z})]] \\
\textrm{subj. to} \quad D_{KL}[q_\phi(\mathbf{z} \mid X) || p_\theta(\mathbf{z} \mid Y)] < \epsilon
\end{align}
}
Here $D$ denotes the dataset of both images $X$ and segmentations $Y$, the objective function aims to maximise the probability of the observed segmentation $Y$ given the latent space distribution $q_\phi(\mathbf{z} \mid X)$. A Kullback-Leibler (KL) divergence $D_{KL}$ is introduced to match the distribution $q_\phi(\mathbf{z} \mid X)$ and $p_\theta(\mathbf{z} \mid Y)$.

Inspired by $\beta$-VAE \cite{Higgins2017betaVAELB} and rewriting the optimisation problem using the Lagrangian multiplier $\beta$, we have,
\begin{align}
	\mathcal{F}(\theta, \phi, \beta;Y,\mathbf{z}) & = \mathbb{E}_{q_\phi(\mathbf{z} \mid X)}[\log p_\theta(Y \mid \mathbf{z})] - \beta (D_{KL}[q_\phi(\mathbf{z} \mid X) || p_\theta(\mathbf{z} \mid Y)] - \epsilon) \\
	\label{elbo}
    & \geq \mathbb{E}_{q_\phi(\mathbf{z} \mid X)}[\log p_\theta(Y \mid \mathbf{z})] - \beta D_{KL}[q_\phi(\mathbf{z} \mid X) || p_\theta(\mathbf{z} \mid Y)]
\end{align}

We aim to maximise the lower bound of the objective function,
\begin{equation}
	\mathcal{L} = \mathbb{E}_{q_\phi(\mathbf{z} \mid X)}[\log p_\theta(Y \mid \mathbf{z})] - \beta D_{KL}[q_\phi(\mathbf{z} \mid X) || p_\theta(\mathbf{z} \mid Y)]
\end{equation}
In this work, we employ the binary cross entropy loss for the first term, which evaluates the reconstruction accuracy for the predicted segmentation, compared to ground truth. $\beta$ denotes a regularization weight, which encourages the posterior $q_\phi$ to be close to the prior $p_\theta$. 

\subsection{Hierarchical Representation}
Without loss of generality, we assume a multi-level network such as U-net is used for the segmentation task, which contains skip connections at multiple resolution levels. These skip connections encode rich image features, which allow us to model the data uncertainties at different resolution levels. Given a U-net of $L+1$ resolution levels, we introduce a number of latent variables $z_i (i = 0, \dots, L)$ to describe the latent representation at resolution level $i$. These latents are inferred from the feature maps at the skip connections. They are inter-dependent as fine resolution features are normally built upon coarse resolution features. For instance, $z_0$ denotes the latent variable from the top level skip connection features extracted directly from the input image. $z_1$ denotes the latent variable from the second level, which has a dependency on $z_0$. Therefore, the image-to-latent encoder $q_\phi$ and segmentation-to-latent encoder $p_\theta$ can be expanded with the product rule:
\begin{align}
	q_\phi(\mathbf{z} \mid X) & = q_\phi(z_L \mid z_{<L}, X)\cdot...\cdot q_\phi(z_0 \mid X) \\
	p_\theta(\mathbf{z} \mid Y) & = p_\theta(z_L \mid z_{<L}, Y)\cdot...\cdot p_\theta(z_0 \mid Y)
\end{align}

The KL-divergence between the two distribution can be expanded hierarchically \cite{DBLP:journals/corr/abs-1905-13077} as, 

\begin{align}
	&\quad D_{KL}[q_\phi(\mathbf{z} \mid X) || p_\theta(\mathbf{z} \mid Y)]\\ &= \mathbb{E}_{z\sim q_\phi(\mathbf{z} \mid X)} [\log q_\phi(\mathbf{z} \mid X) - \log p_\theta(\mathbf{z} \mid Y)] \\
	&= \int_{\mathbf{z}}\Pi^L_{j = 0}q_\phi(z_j \mid z_{<j}, X)\sum^L_{i = 0}[\log q_\phi(z_i \mid z_{<i}, X) - \log p_\theta(z_i \mid z_{<i}, Y)] dz_0...dz_L\\
	&= \sum^L_{i = 0}\int_{\mathbf{z}}\Pi^i_{j = 0}q_\phi(z_j \mid z_{<j}, X)[\log q_\phi(z_i \mid z_{<i}, X) - \log p_\theta(z_i \mid z_{<i}, Y)] dz_0...dz_L\\
	&= \sum^L_{i = 0}\int_{\mathbf{z}}\Pi^{i-1}_{j = 0}q_\phi(z_j \mid z_{<j}, X)q_\phi(z_i \mid z_{<i}, X)[\log q_\phi(z_i \mid z_{<i}, X) \\&\qquad- \log p_\theta(z_i \mid z_{<i}, Y)] dz_0...dz_L\\
	&= \sum^L_{i = 0}\mathbb{E}_{\mathbf{z}_{<i}\sim q_\phi(\mathbf{z} \mid X)} [D_{KL}[q_\phi(z_{i} \mid z_{<i}, X) || p_\theta(z_{i} \mid z_{<i}, Y)]]
 \label{eq:hkl}
\end{align}
This means that the KL-divergence term can be evaluated separately for each resolution level, which eases the implementation. Note that for simplicity, $p_\theta(z_0\mid z_{<0})$ and $q_\phi(z_0\mid z_{<0})$ represents for $p_\theta(z_0)$ and $q_\phi(z_0)$ separately. Plugging this formulation into the KL-divergence in Eq. (\ref{eq:hkl}), we have the following objective function,
\begin{equation}
	\mathcal{L} = \mathbb{E}_{q_\phi(\mathbf{z} \mid X)}[\log p_\theta(Y \mid \mathbf{z})] - \beta \sum^{L}_{i = 0}\mathbb{E}_{\mathbf{z}_{<i}\sim q_\phi(\mathbf{z} \mid X)} [D_{KL}[q_\phi(z_{i} \mid z_{<i}, X) || p_\theta(z_{i} \mid z_{<i}, Y)]]
\end{equation}
The expectation of the first term can be performed with a single pass Monte-Carlo sampling \cite{DBLP:journals/corr/abs-1906-02691} in the implementation.

\subsection{Proposed Architecture}
\begin{figure}[!h]
    \centering
    \includegraphics[width=12cm]{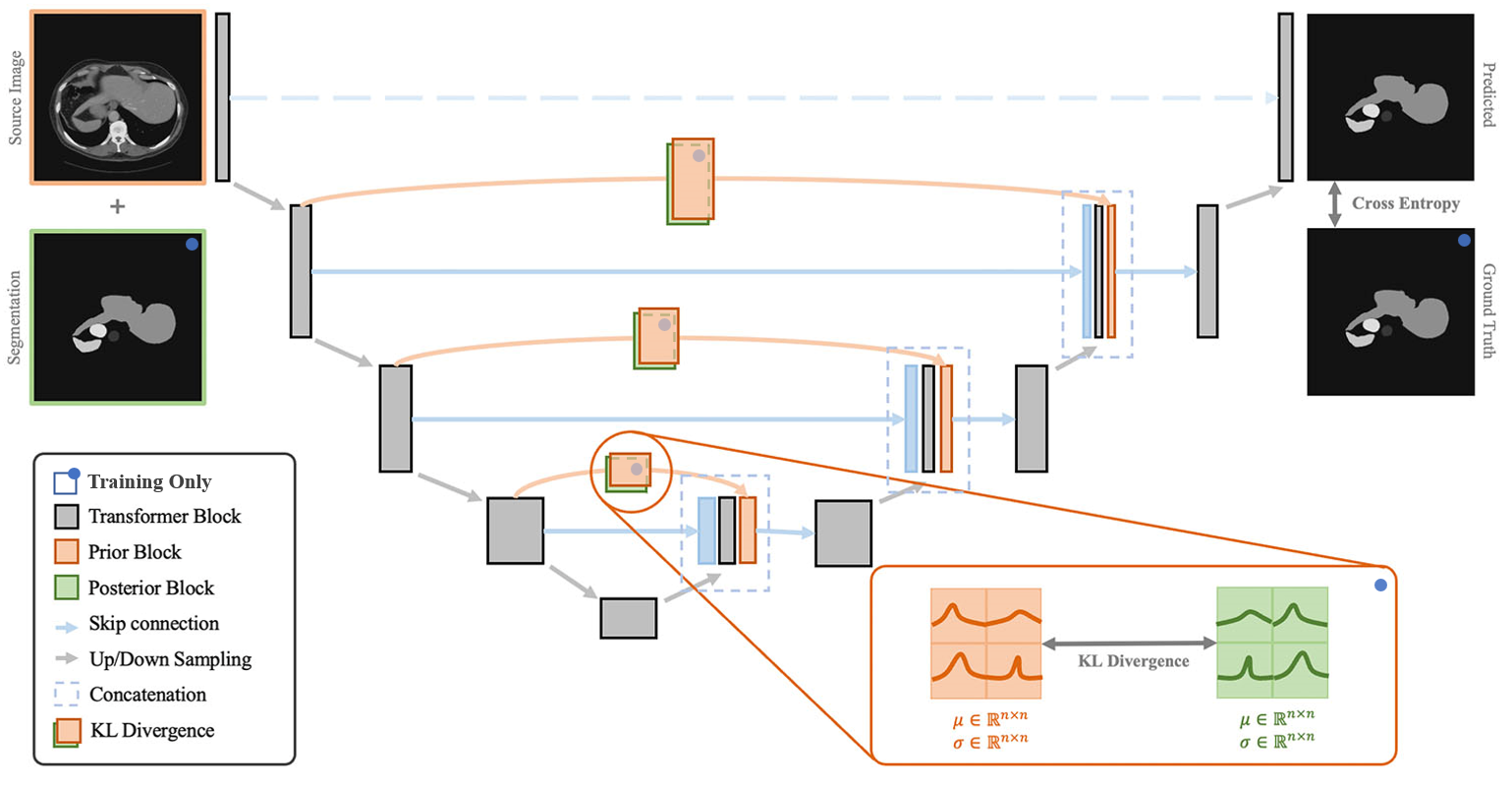}
    \caption{Hierarchical uncertainty estimation architecture. The structure of the network is based on U-net, with VAE modules that accompany the skip connections to model the uncertainty at multiple resolution levels.}
    \label{fig:training}
\end{figure}

Figure \ref{fig:training} illustrates the overall structure of the proposed segmentation framework with uncertainty awareness, named as VAE U-net. 

The network follows the typical setup of a U-net architecture, composed of encoder and decoder modules. At the training stage, it takes both the image $X$ and the segmentation $Y$, aiming to generate the correct segmentation while at the meantime matching the latent distributions $q_\phi(\mathbf{z} \mid X)$ and $p_\theta(\mathbf{z} \mid Y)$. The latents are modelled along the skip connections at each resolution level. At level $i$, 1$\times$1 convolutions are applied to the skip connection feature map, which generate pixel-wise parameter maps for the latents, $\boldsymbol{\mu}_i^{\text {prior }}\left(\mathbf{z}_{<i}, X\right)$, $\boldsymbol{\sigma}_i^{\text {prior }}\left(\mathbf{z}_{<i}, X\right)$ for image latents, and $\boldsymbol{\mu}_i^{\text {prior }}\left(\mathbf{z}_{<i}, Y\right)$, $\boldsymbol{\sigma}_i^{\text {prior }}\left(\mathbf{z}_{<i}, Y\right)$ for segmentation latents. The latents at level $i$ are sampled from the Gaussian distribution,
\begin{equation}
    \mathbf{z}_i \sim \mathcal{N}\left(\boldsymbol{\mu}_i^{\text {prior }}\left(\mathbf{z}_{<i}, X\right), \boldsymbol{\sigma}_i^{\text {prior }}\left(\mathbf{z}_{<i}, X\right)\right)=: p_\theta\left(\mathbf{z}_i \mid \mathbf{z}_{<i}, X\right)
\end{equation}
and then concatenated with other feature maps and passed on to the U-net decoder. The mean $\boldsymbol{\mu}_i$ and standard deviation $\boldsymbol{\sigma}_i$ model the uncertainty at this level. Once the model is trained, at the inference stage, the decoder samples from the image latent space and reconstructs the segmentation map. Comparing to previous works \cite{PhiSeg2019,kohl2019hierarchical}, this new design leverages the multi-level feature maps from skip connections and performs sampling in the latent space before the decoder, therefore allows the decoder to learn how to interpret the latent space.

\section{Experiments}

\subsection{Setup}
Within this work, we aim to improve both the uncertainty estimation as well as the segmentation performance. To this end, for the first part, we evaluate the uncertainty estimation ability quantitatively using the LIDC-IDRI dataset (The Lung Image Database Consortium image collection \cite{armato2011lung}, licensed under \href{https://wiki.cancerimagingarchive.net/x/c4hF}{TCIA Data Usage Policy and Restrictions}) and compare to other uncertainty-aware methods including hierarchical probabilistic U-net \cite{kohl2019hierarchical} and PHiSeg \cite{PhiSeg2019}. Following up as the second part, we also perform quantitative analysis with manually crafted out-of-distribution samples to show the usefulness of the uncertainty map. For the last part, we conduct experiments on Synapse dataset (a multi-organ segmentation dataset \cite{synapse} contains 30 abdominal CT scans with 3,779 axial clinical CT images, licensed under \href{https://creativecommons.org/licenses/by/4.0/legalcode}{CC BY 4.0}) and compare segmentation quality to state-of-the-art segmentation models, including the Unet \cite{ronneberger2015u}, AttnUnet \cite{AttnUNet}, SwinUnet \cite{cao2021swin} and TransUnet \cite{chen2021transunet}.

\subsection{Results}
\subsubsection{Quantitative Metrics for Uncertainty Modeling}
\begin{figure}[!h]
    \centering
    \includegraphics[width=11cm]{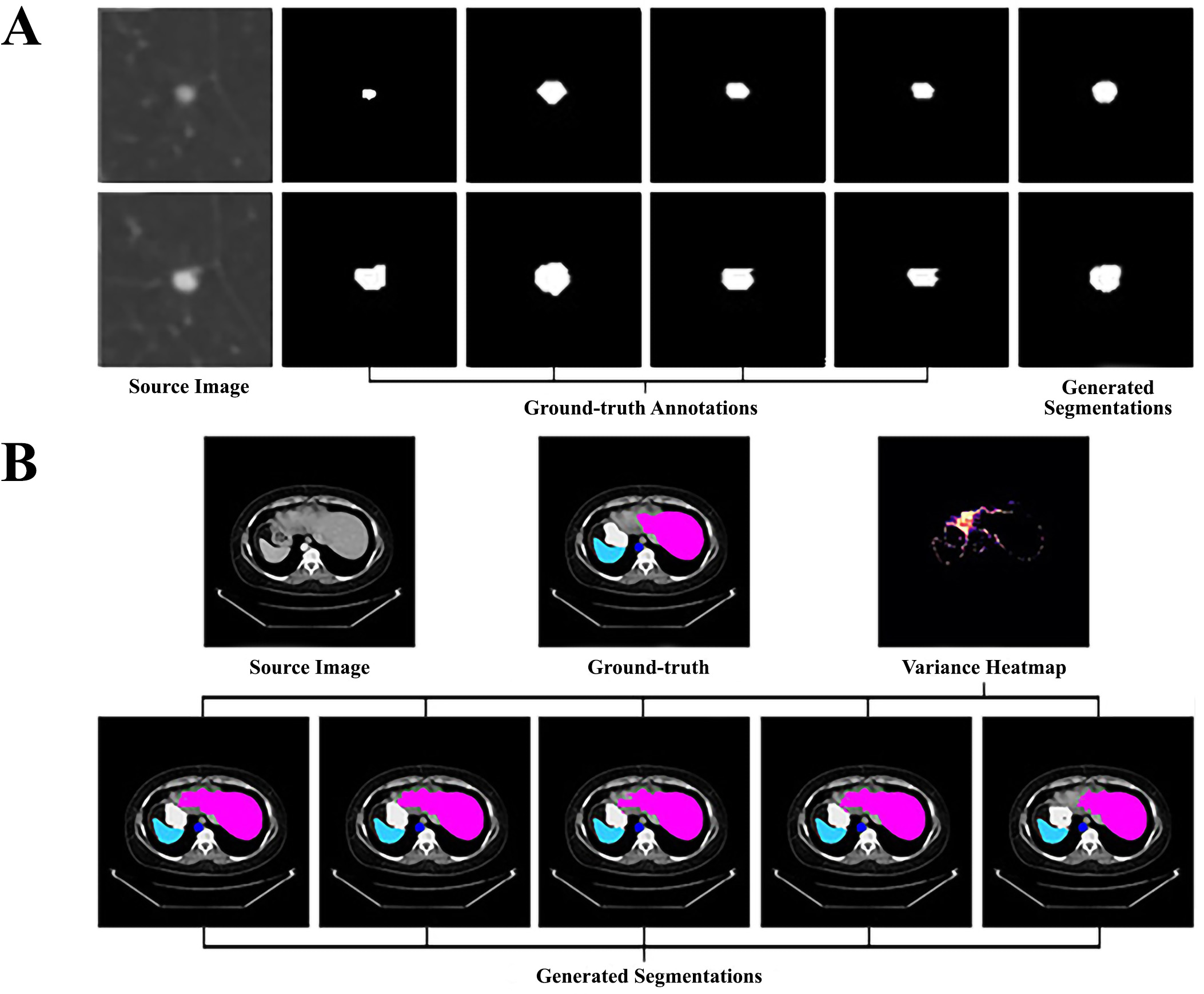}
    \caption{Generative Segmentation of Medical Images with VAE U-net. A) Generated Lung cancer segmentation from  LIDC-IRDI dataset. It is able to provide a plausible boundary based on four ground truth segmentations. B) Segmentation of abdominal organs on Synapse dataset with the variance heatmap. The variance heatmap is derived from multiple sampled segmentations.}
    \label{fig:sample}
\end{figure}

Following the experiment setup as in previous work \cite{kohl2019hierarchical}, we trained the proposed model, VAE U-net, using images and paired annotation from only one expert. Then the model was evaluated on a test subset, where all four experts agreed on the presence of the lung lesion but provided independent annotations. On this test subset, automated lung lesion segmentations were generated by drawing 100 random samples from the latent space. Following \cite{PhiSeg2019}, two metrics, $D_{\mathrm{GED}}^2$ and $\mathcal{S}_{\mathrm{NCC}}$, were evaluated, which compare the automated segmentations to multiple expert annnotations. The generalized energy distance $D_{\mathrm{GED}}^2$ is defined as $
    D_{\mathrm{GED}}^2\left(p_{t}, p_{\mathbf{s}}\right)=2 \mathbb{E}[d(\mathbf{s}, \mathbf{t})]-\mathbb{E}\left[d\left(\mathbf{s}, \mathbf{s}^{\prime}\right)\right]-\mathbb{E}\left[d\left(\mathbf{t}, \mathbf{t}^{\prime}\right)\right]
$
where $d(*, *)=1-\operatorname{IoU}(*, *)$, $s, s^\prime$ are drawn from the learnt distribution $p_{\mathbf{s}}$ and $t, t^\prime$ are chosen from the ground-truth annotation $p_{\mathbf{t}}$. The first term measures the similarity between the segmentation and the ground-truth annotation, whereas the second and the last terms measure the variability of the predicted segmentations and the ground-truth annotations respectively. Generalized energy distance models the resemblance of two distributions on the global manner, but may not be sensitive to details. The normalized cross correlation score $\mathcal{S}_{\mathrm{NCC}}$ is used to evaluate the distribution similarity at the pixel level. It is defined as, $
\mathcal{S}_{\mathrm{NCC}}\left(p_{t}, p_{\mathbf{s}}\right)=\mathbb{E}_{\mathbf{t} \sim p_{t}}\left[\mathrm{NCC}\left(\mathbb{E}_{\mathbf{s} \sim p_{\mathbf{s}}}[\mathrm{CE}(\overline{\mathbf{s}}, \mathbf{s})], \mathbb{E}_{\mathbf{s} \sim p_{\mathbf{s}}}[\mathrm{CE}(\mathbf{t}, \mathbf{s})]\right)\right]
$, where $CE(*, *)$ stands for cross-entropy.
\begin{table}[!h]

\center
\setlength{\tabcolsep}{0.4mm}{
\begin{tabular}{cccc}
\hline
Methods & GED $\downarrow$ & NCC $\uparrow$ & Parameters $\downarrow$     \\
\hline
H.Prob. U-net \cite{kohl2019hierarchical} & 0.4452 & 0.5999 & $\mathbf{5.00}$ M \\
PHiSeg (L=1) \cite{PhiSeg2019} & 0.4695 & 0.6013 & 26.37 M \\
PHiSeg (L=5) \cite{PhiSeg2019} & \underline{0.3225} & $\mathbf{0.7337}$ & 54.37 M \\
\hline
VAE U-net (Proposed) & $\mathbf{0.2888}$ & \underline{0.7087} & $\underline{20.08}$ M \\ 
\hline
\end{tabular}
}
\newline
\caption{Evaluation on LIDC test subset. The metrics are calculated from 100 generated segmentations sampled from prior latent distribution. The highest-performing metrics are presented in bold, while the second-place metric are underscored.}
\label{table:lidc:new}
\end{table}
\vspace{-2em}
Table \ref{table:lidc:new} reports the GED, NCC as well as the number of model parameters (in million), comparing the proposed method to other methods that account for segmentation uncertainties. The proposed method outperforms all other uncertainty-aware methods for GED metrics, while performing comparatively well in NCC with PHiSeg (L=5) but with fewer parameters. This shows that the distribution of the generated segmentations by the proposed method agrees well with the ground truth distribution given by four experts, even when trained on single annotations only. The top panel of Figure \ref{fig:sample} provides some visualization of the generated lesion segmentations.

\subsubsection{Qualitative Study of the Uncertainty Map}
\begin{figure}[h]
    \centering
    \includegraphics[width=6cm]{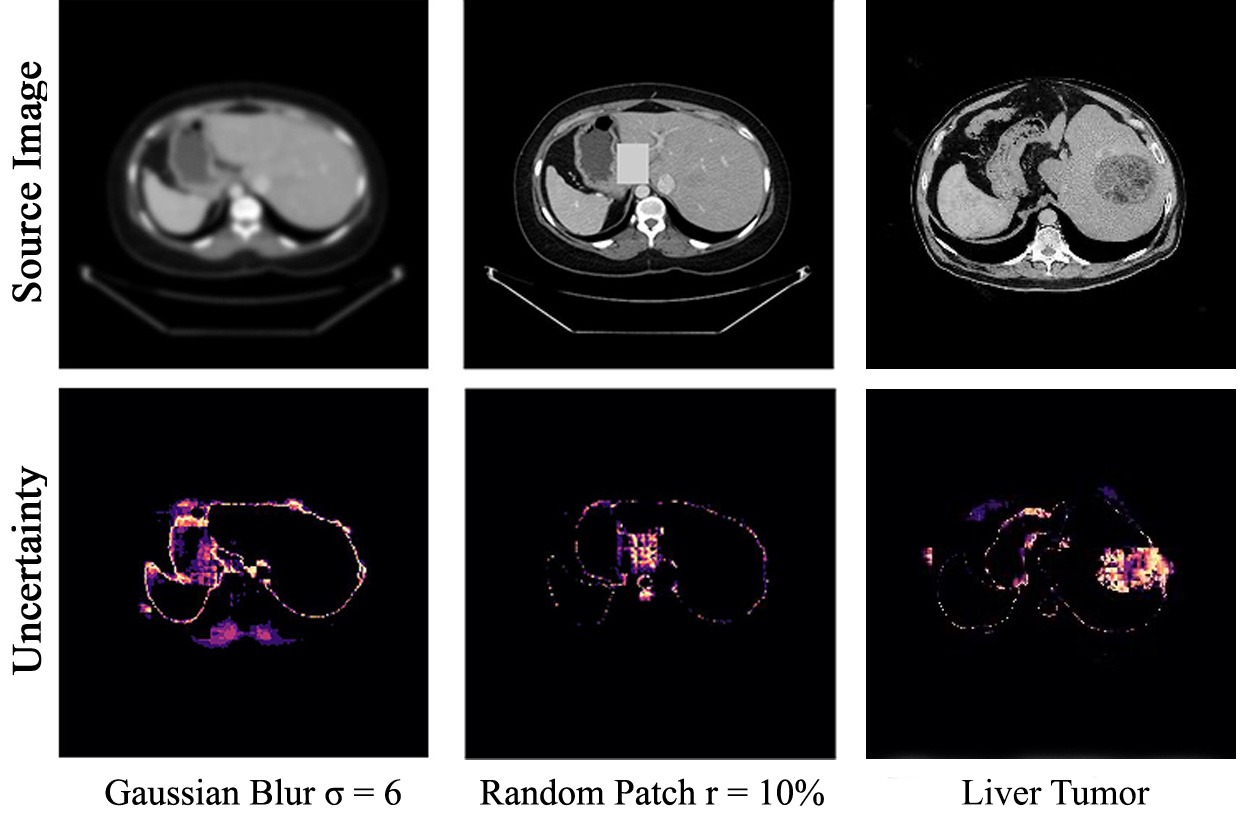}
    \caption{Uncertainty heatmap for out-of-distribution detection. The heatmap is generated from 10 segmentation maps from VAE U-net. Brighter color means more disagreement (variance) in generated segmenatations and vice versa. Random patching adds a $r=10\%$ patch which is proportional to 10\% of the image size.}
    \label{fig:out-of-distribution}
\end{figure}

After drawing multiple segmentations from the latent space using VAE U-net, we can generate a segmentation uncertainty map by calculating the pixel-wise variance of predicted segmentations and normalizing it by the maximum value of variance among the samples, illustrated by the bottom panel of Figure \ref{fig:sample}. We demonstrate the use of this uncertainty map for the out-of-distribution detection task. Out-of-distribution samples are constructed using three different ways: 1) applying Gaussian blurring to the image; 2) adding a patch to a random location in the image; 3) using abnormal samples from PMC Open Access Subset \cite{gamble2017pubmed}. Figure \ref{fig:out-of-distribution} visualizes the uncertainty maps for the three cases. For Gaussian blurred images, the map highlights the uncertainty near the anatomical boundaries. For random patched image or abnormal image with liver tumour, the map highlights the location of abnormalities. The ability of uncertainty measurement can serve as a potentially powerful tool for further downstream tasks.

\subsubsection{Quantitative Metrics for Segmentation Performance}
Table \ref{table:synapse} compares the segmentation performance of the proposed method to state-of-the-art deep learning methods on the Synapse dataset. The proposed VAE U-net was used in two modes: the sampling mode which generates a number of plausible segmentations and the prior mode which generates a single segmentation using the prior mean. Both modes show a strong performance compared to existing methods. VAE U-net in prior mode achieves the highest overall Dice score of 78.71\% on Dice score and the highest overall Hausdorff distance (HD) of 20.86mm. It also achieved the best score for the kidney and stomach. In sampling mode, the sampled segmentations achieve $77.29\% (\pm0.22\%)$ in Dice and $26.52\text{mm}(\pm1.65\text{mm})$ in HD, which are in line with other state-of-the-art methods, indicating the usefulness of incorporating uncertainty into segmentation. The bottom panel of Figure \ref{fig:sample} shows the visualization of the segmentations as well as the variance on selected samples.

\begin{table}[h!]

\center
\resizebox{\textwidth}{!}{
\begin{tabular}{c|c|c|c|c|c|c|c|c|c|c}
\hline
Methods                             & Dice $\uparrow$         & HD $\downarrow$          & Aorta        & Gallbladder  & Kidney(L)    & Kidney(R)    & Liver        & Pancreas     & Spleen       & Stomach      \\
\hline
V-Net                               & 68.81        & -            & 75.34        & 51.87        & 77.10        & \textbf{80.75}        & 87.84        & 40.05        & 80.56        & 56.98        \\
DARR                                & 69.77        & -            & 74.74        & 53.77        & 72.31        & 73.24        & \textbf{94.08}        & 54.18        & \textbf{89.90}        & 45.96        \\
R50 U-Net                           & 74.68        & 36.87        & 87.74        & 63.66        & 80.60        & 78.19        & 93.74        & 56.90        & 85.87        & 74.16        \\
U-Net                               & 76.85        & 39.70        & 89.07        & \textbf{69.72}        & 77.77        & 68.60        & 93.43        & 53.98        & 86.67        & 75.58        \\
R50 Att-Unet                        & 75.57        & 36.97        & 55.92        & 63.91        & 79.20        & 72.71        & 93.56        & 49.37        & 87.19        & 74.95        \\
Att-Unet                            & 77.77        & 36.02        & \textbf{89.55}        & 68.88        & 77.98        & 71.11        & 93.57        & \textbf{58.04}        & 87.30        & 75.75        \\
R50 ViT                             & 71.29        & 32.87        & 73.73        & 55.13        & 75.80        & 72.20        & 91.51        & 45.99        & 81.99        & 73.95        \\
TransUnet                           & 77.48        & 31.69        & 87.23        & 63.13        & 81.87        & 77.02        & 94.08        & 55.86        & 85.08        & 75.62        \\
SwinUnet                            & 78.39        & 24.73        & 86.65        & 66.15        & 83.71        & 79.89        & 93.87        & 56.00        & 87.98        & 72.88        \\
\hline
\makecell{VAE U-net (Sample)} & \makecell{77.29 \\$\pm$ 0.22} & \makecell{26.52 \\$\pm$ 1.65} & \makecell{83.99 \\$\pm$ 0.22} & \makecell{63.37 \\$\pm$ 0.74} & \makecell{82.93 \\$\pm$ 0.49} & \makecell{78.42 \\$\pm$ 0.46} & \makecell{93.59 \\$\pm$ 0.06} & \makecell{53.38 \\$\pm$ 0.55} & \makecell{88.78 \\$\pm$ 0.30} & \makecell{73.83 \\$\pm$ 0.41} \\
\makecell{VAE U-net (Prior)}  & \textbf{78.71}        & \textbf{20.86}        & 85.40        & 65.01        & \textbf{83.96}        & 80.16        & 93.89        & 55.42        & 89.65        & \textbf{76.18} \\ 
\hline
\end{tabular}
}
\newline
\caption{Segmentation performance on Synapse test set. The second and third columns report the average performance across organs. The mean and variance in sampling mode is calculated from 10 generated segmentations sampled from the prior latent distribution.}
\label{table:synapse}
\end{table}

\section{Conclusion}
In this paper, we proposed a novel uncertainty measurement method for Unet-based image segmentation models, named as VAE U-net. VAE U-net models uncertainty via the skip connection and thus is potentially applicable to other segmentation models that consist of multi-level skip connections. The proposed method achieves a high segmentation performance compared to state-of-the-art deterministic models, as well as to a high uncertainty estimation capability compared to uncertainty-aware segmentation models in terms of generalised energy distance and normalized cross correlation score. By drawing multiple plausible segmentations from the latent space, the proposed method can generate an uncertainty map for the segmentation, which not only facilitates the detection of out-of-distribution samples with noise and abnormality, but also provide an extra layer of interpretability to the segmentation.

\newpage
\bibliographystyle{splncs04}
\bibliography{bibliography}
\newpage

%
%
%

\section*{Appendix}
\appendix
\section{Model Output and Visualization for Synapse Dataset}

In the paper, we quantitively evaluate model performance with the sampled latent space with the prior distribution. The detailed metrics for each sample are shown in Table \ref{table:variance}.

\begin{table}[htbp]
\center
\resizebox{\textwidth}{!}{
\begin{tabular}{c|c|c|c|c|c|c|c|c|c|c|c|c|c|c|c|c}
\hline
Sample & \multicolumn{2}{c}{Aorta}         & \multicolumn{2}{c}{Gallbladder}         & \multicolumn{2}{c}{Kidney(L)}         & \multicolumn{2}{c}{Kidney(R)}         & \multicolumn{2}{c}{Liver}        & \multicolumn{2}{c}{Pancreas}        & \multicolumn{2}{c}{Spleen}         & \multicolumn{2}{c}{Stomach}        \\
 & DSC    & HD     & DSC     & HD          & DSC     & HD        & DSC     & HD        & DSC     & HD     & DSC     & HD       & DSC     & HD     & DSC     & HD \\
\hline
1      & 0.8366 & 10.8351 & 0.6369      & 42.0419 & 0.8309    & 35.9437 & 0.7920    & 48.9854 & 0.9356 & 21.1717 & 0.5280   & 13.7971 & 0.8883 & 30.1600 & 0.7416  & 18.1989 \\
2      & 0.8447 & 7.5771  & 0.6371      & 28.9042 & 0.8297    & 32.5005 & 0.7843    & 56.9210 & 0.9360 & 21.3996 & 0.5283   & 13.3646 & 0.8888 & 36.6232 & 0.7423  & 16.8864 \\
3      & 0.8394 & 9.4634  & 0.6377      & 36.7390 & 0.8267    & 32.9405 & 0.7835    & 48.2171 & 0.9364 & 20.1728 & 0.5394   & 13.9482 & 0.8872 & 32.2708 & 0.7366  & 19.1086 \\
4      & 0.8370 & 10.6003 & 0.6416      & 34.6728 & 0.8283    & 33.6209 & 0.7863    & 51.2560 & 0.9364 & 20.6927 & 0.5250   & 14.1900 & 0.8847 & 24.5469 & 0.7367  & 17.6837 \\
5      & 0.8422 & 7.9080  & 0.6394      & 37.1049 & 0.8298    & 39.9077 & 0.7913    & 44.3016 & 0.9358 & 20.9809 & 0.5303   & 13.9811 & 0.8908 & 23.4351 & 0.7422  & 17.6694 \\
6      & 0.8392 & 11.5661 & 0.6457      & 42.0551 & 0.8313    & 43.2612 & 0.7757    & 52.3549 & 0.9361 & 20.7980 & 0.5281   & 14.0435 & 0.8911 & 21.3762 & 0.7429  & 16.8955 \\
7      & 0.8360 & 10.1318 & 0.6150      & 43.8266 & 0.8227    & 38.1155 & 0.7798    & 51.0938 & 0.9349 & 20.8692 & 0.5291   & 13.3467 & 0.8876 & 33.1373 & 0.7324  & 17.3859 \\
8      & 0.8416 & 9.3573  & 0.6268      & 35.4166 & 0.8343    & 32.0674 & 0.7827    & 37.2433 & 0.9352 & 22.7996 & 0.5427   & 13.4555 & 0.8865 & 26.1281 & 0.7384  & 17.6222 \\
9      & 0.8384 & 9.9281  & 0.6243      & 43.1226 & 0.8347    & 31.3610 & 0.7891    & 37.8275 & 0.9370 & 21.1330 & 0.5359   & 12.9866 & 0.8919 & 23.7895 & 0.7441  & 17.2179 \\
10     & 0.8388 & 8.8284  & 0.6314      & 36.4831 & 0.8295    & 35.0589 & 0.7807    & 43.4456 & 0.9359 & 26.7443 & 0.5423   & 13.3481 & 0.8816 & 26.4568 & 0.7360  & 16.6224 \\
11     & 0.8393 & 10.1432 & 0.6250      & 44.0193 & 0.8364    & 31.6526 & 0.7892    & 47.2165 & 0.9353 & 20.9640 & 0.5378   & 13.2315 & 0.8864 & 37.1440 & 0.7407  & 17.0977 \\
12     & 0.8377 & 10.4850 & 0.6367      & 38.5701 & 0.8353    & 35.5516 & 0.7871    & 54.8652 & 0.9352 & 22.0824 & 0.5358   & 13.3616 & 0.8861 & 26.5290 & 0.7395  & 17.3862 \\
13     & 0.8401 & 7.5355  & 0.6306      & 43.4111 & 0.8243    & 32.4223 & 0.7800    & 60.8630 & 0.9359 & 20.2417 & 0.5349   & 13.8721 & 0.8865 & 22.4895 & 0.7266  & 18.0815 \\
14     & 0.8377 & 9.5309  & 0.6326      & 29.2546 & 0.8232    & 41.4945 & 0.7823    & 54.1239 & 0.9365 & 20.9434 & 0.5293   & 13.7607 & 0.8911 & 19.2319 & 0.7371  & 16.6482 \\
15     & 0.8408 & 12.2123 & 0.6317      & 34.6952 & 0.8319    & 33.1501 & 0.7892    & 44.0921 & 0.9370 & 20.6693 & 0.5360   & 14.1194 & 0.8920 & 25.7182 & 0.7381  & 17.2505 \\
16     & 0.8390 & 10.3945 & 0.6285      & 35.8092 & 0.8289    & 36.4739 & 0.7872    & 43.1464 & 0.9347 & 21.9926 & 0.5276   & 13.7141 & 0.8849 & 25.9178 & 0.7311  & 18.3067 \\
17     & 0.8433 & 7.7038  & 0.6214      & 43.8646 & 0.8323    & 32.6734 & 0.7884    & 50.8036 & 0.9366 & 20.8931 & 0.5386   & 14.2091 & 0.8855 & 23.8639 & 0.7431  & 16.0795 \\
18     & 0.8397 & 13.0554 & 0.6484      & 41.6519 & 0.8238    & 36.3828 & 0.7710    & 70.8122 & 0.9357 & 25.2957 & 0.5278   & 13.5820 & 0.8854 & 28.7111 & 0.7381  & 18.1320 \\
19     & 0.8397 & 10.1656 & 0.6290      & 35.6434 & 0.8227    & 32.5431 & 0.7767    & 68.4746 & 0.9358 & 21.8821 & 0.5381   & 13.7800 & 0.8804 & 26.7668 & 0.7345  & 17.3851 \\
20     & 0.8444 & 6.8438  & 0.6415      & 35.0602 & 0.8349    & 31.7220 & 0.7882    & 42.0353 & 0.9362 & 21.2437 & 0.5398   & 13.7009 & 0.8895 & 24.7836 & 0.7402  & 19.3254 \\
21     & 0.8402 & 9.3683  & 0.6403      & 35.1964 & 0.8324    & 31.6422 & 0.7856    & 42.2339 & 0.9351 & 33.4777 & 0.5363   & 14.0311 & 0.8868 & 29.1294 & 0.7448  & 17.4502 \\
22     & 0.8415 & 8.1393  & 0.6246      & 36.0243 & 0.8388    & 33.4439 & 0.7856    & 35.7542 & 0.9353 & 24.3854 & 0.5323   & 14.5034 & 0.8913 & 25.0883 & 0.7415  & 17.0907 \\
23     & 0.8398 & 9.8867  & 0.6386      & 43.7625 & 0.8364    & 31.5464 & 0.7876    & 48.1456 & 0.9361 & 22.9412 & 0.5437   & 13.6355 & 0.8919 & 24.6596 & 0.7403  & 17.5962 \\
24     & 0.8400 & 8.9493  & 0.6377      & 43.7889 & 0.8291    & 34.2538 & 0.7841    & 35.4504 & 0.9368 & 22.6263 & 0.5375   & 13.9456 & 0.8909 & 22.9949 & 0.7377  & 17.7307 \\
25     & 0.8379 & 11.1249 & 0.6262      & 43.8527 & 0.8248    & 37.3052 & 0.7816    & 63.9425 & 0.9365 & 20.5847 & 0.5302   & 13.9278 & 0.8861 & 26.2930 & 0.7401  & 18.0096 \\
26     & 0.8411 & 11.7145 & 0.6366      & 35.3857 & 0.8263    & 33.3029 & 0.7843    & 27.9618 & 0.9358 & 23.4411 & 0.5207   & 14.4669 & 0.8864 & 24.8067 & 0.7397  & 16.8104 \\
27     & 0.8428 & 10.4767 & 0.6346      & 37.0661 & 0.8254    & 43.4891 & 0.7809    & 45.0421 & 0.9350 & 27.0051 & 0.5361   & 13.4993 & 0.8851 & 33.3704 & 0.7334  & 16.8366 \\
28     & 0.8417 & 8.2642  & 0.6332      & 35.6866 & 0.8173    & 41.7104 & 0.7804    & 47.2405 & 0.9365 & 20.7961 & 0.5332   & 13.8991 & 0.8879 & 25.6882 & 0.7324  & 17.0354 \\
29     & 0.8403 & 7.7829  & 0.6367      & 30.3253 & 0.8267    & 40.9993 & 0.7853    & 47.4382 & 0.9355 & 20.1965 & 0.5322   & 14.0804 & 0.8911 & 24.6459 & 0.7390  & 17.1018 \\
30     & 0.8374 & 10.8176 & 0.6419      & 36.5570 & 0.8289    & 44.0330 & 0.7845    & 55.7831 & 0.9358 & 22.0614 & 0.5370   & 14.5356 & 0.8897 & 22.5841 & 0.7389  & 16.9006\\
\hline
\end{tabular}
}
\caption{Model Performance Metrics for 30 Sampled Annotations using Prior Distribution on Synapse Dataset.}
\label{table:variance}
\end{table}

\begin{figure}[H]
    \centering
    \includegraphics[width=12cm]{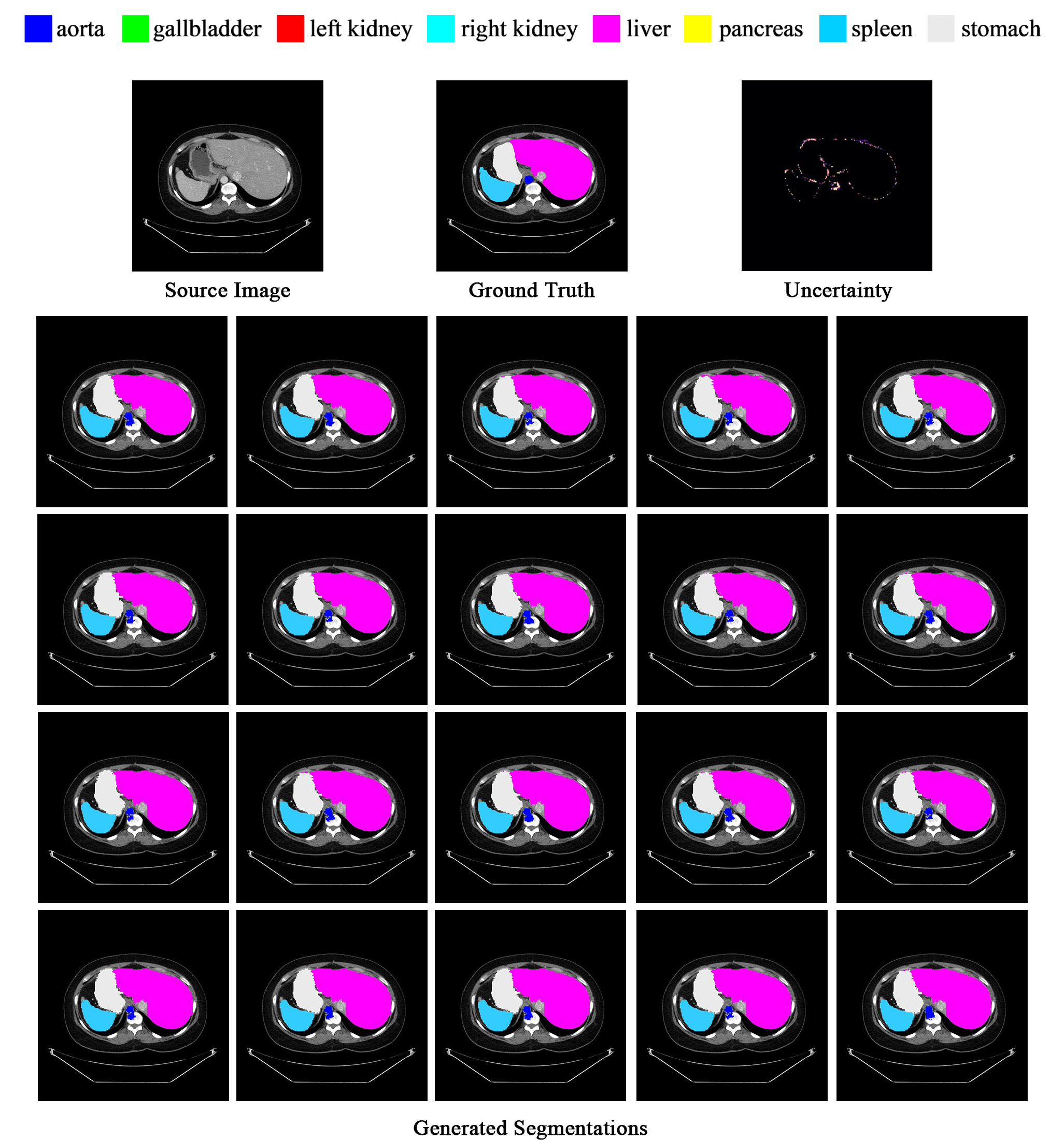}
    \caption{Visualization of Generated Samples with Uncertainty on Synapse dataset Test Case 35 (Slice 80).}
    \label{fig:low-ambiguity}
\end{figure}

In Synapse dataset, the degree of uncertainty largely depends on the slice position. When the slices are closer to the boundary of the organs, the model becomes uncertain about the correct segmentation of the source image. Figure \ref{fig:low-ambiguity} is the generated segmentations for test case 35 (slice 80). We can see the boundary of the organs in the uncertainty magma colour map, which is the variance of the segmentation map (20 in total). In this case, there is little ambiguity in the target segmentation, therefore the generated segmentations look similar.

\begin{figure}[H]
    \centering
    \includegraphics[width=12cm]{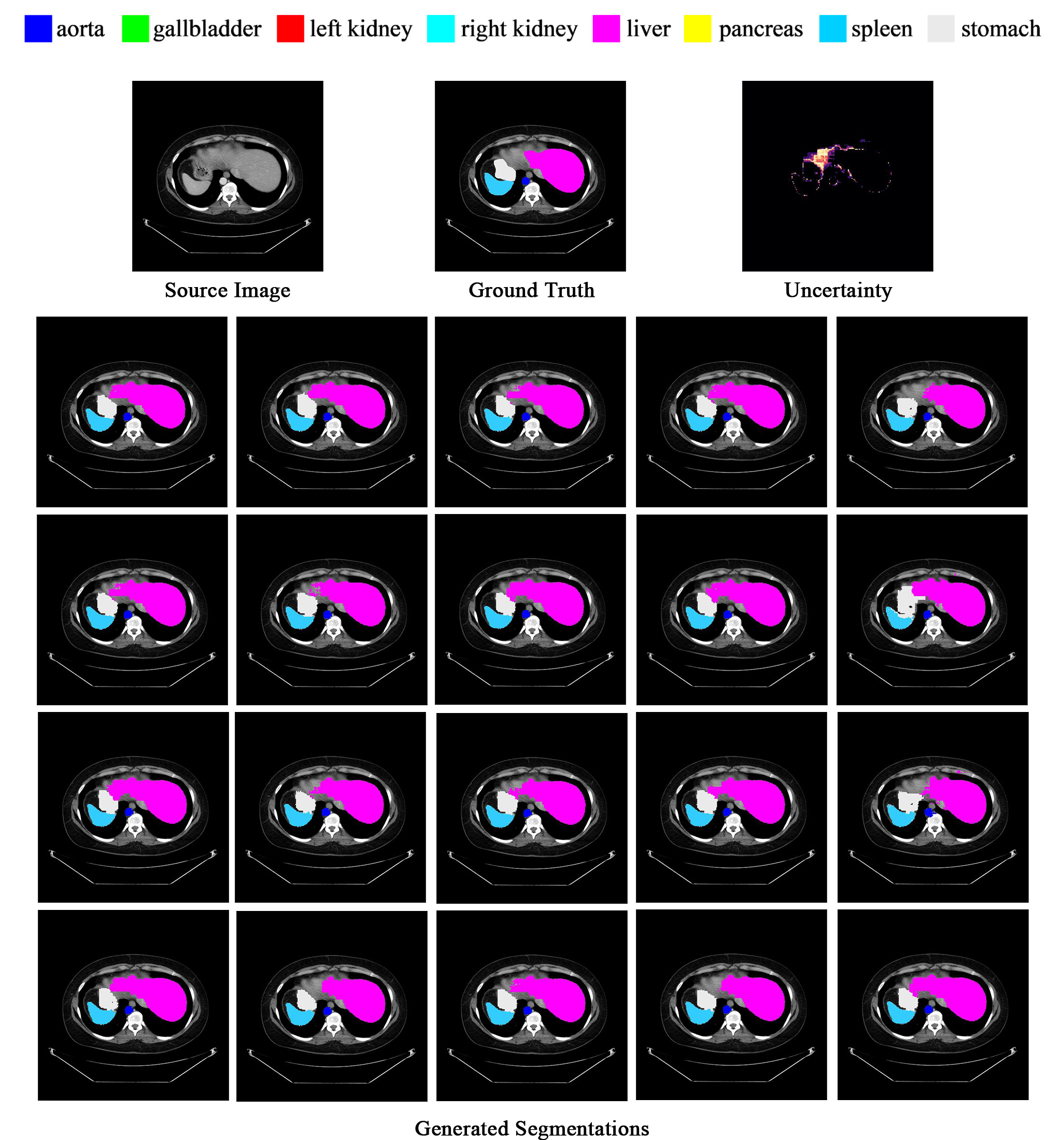}
    \caption{Visualization of Generated Samples with Uncertainty on Synapse dataset Test Case 35 (Slice 84).}
    \label{fig:high-ambiguity}
\end{figure}

For the same test case but with a different slice position, the ambiguity increases as the boundaries between organs become blurry. For slice 84, the visualized samples are shown in Figure \ref{fig:high-ambiguity}. It is clear that the area around the liver in the source image stops the human grader to mark it precisely, but the model can generate different reasonable samples in the learning process and produces high uncertainty measurements around this area.

\begin{figure}[H]
    \centering
    \includegraphics[width=12cm]{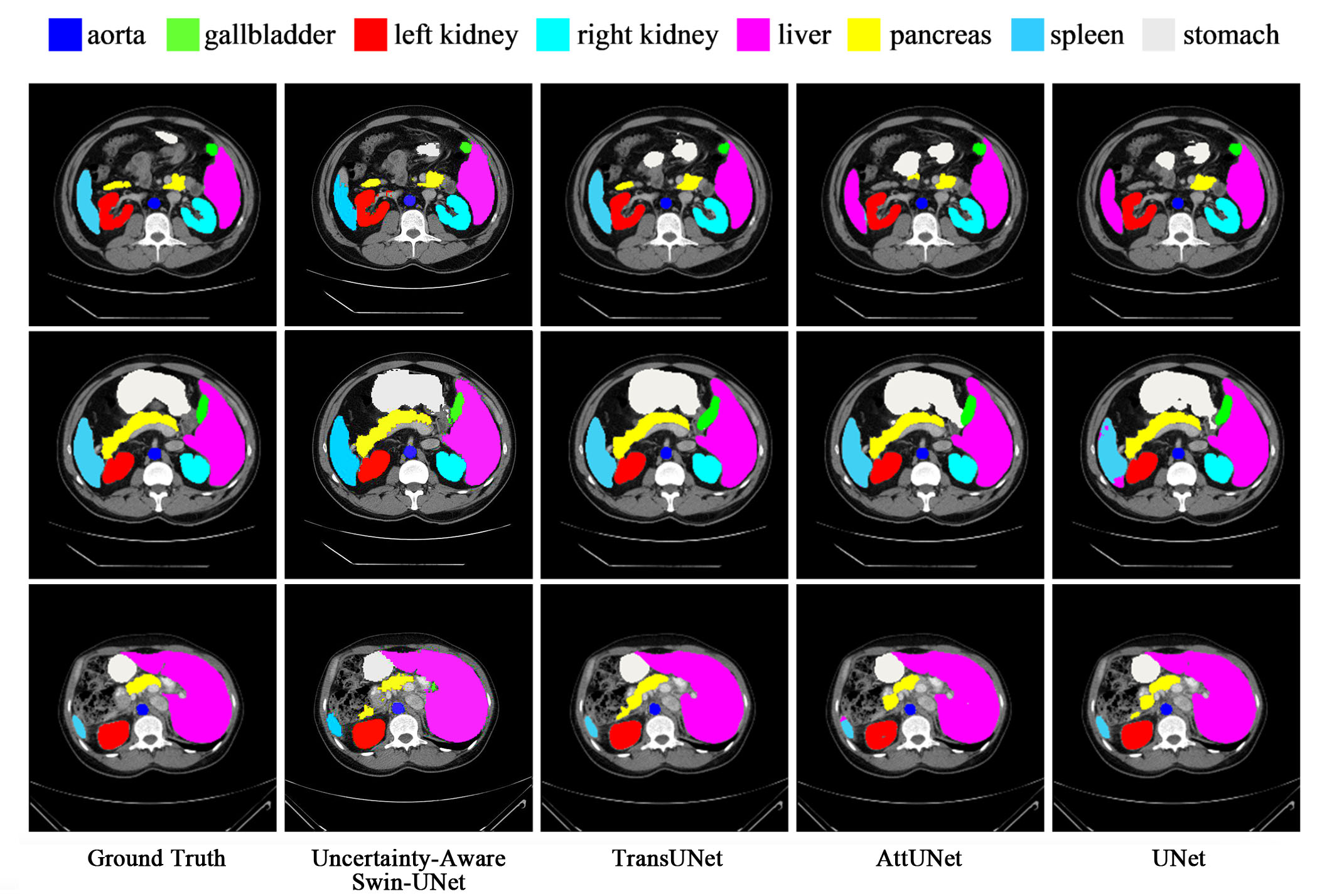}
    \caption{Comparison of Segmentation Results on Synapse Dataset. Models for comparison are TransUnet \cite{chen2021transunet}, AttnUnet \cite{AttnUNet} and Unet \cite{ronneberger2015u}.}
    \label{fig:comparison}
\end{figure}

To comprehensively compare the fidelity of the reconstruction, we picked three previous state-of-the-art Unet-like models to compare against our approach. Three ambiguous segmentation samples are selected from test case 22 (slice 61), test case 22 (slice 66) and test case 2 (slice 106) to visualize the overlaid segmentation map. The result is shown in Figure \ref{fig:comparison}.

\section{Model Prediction Visualization for LIDC Dataset}

\begin{figure}[H]
\centering
    \includegraphics[width=12cm]{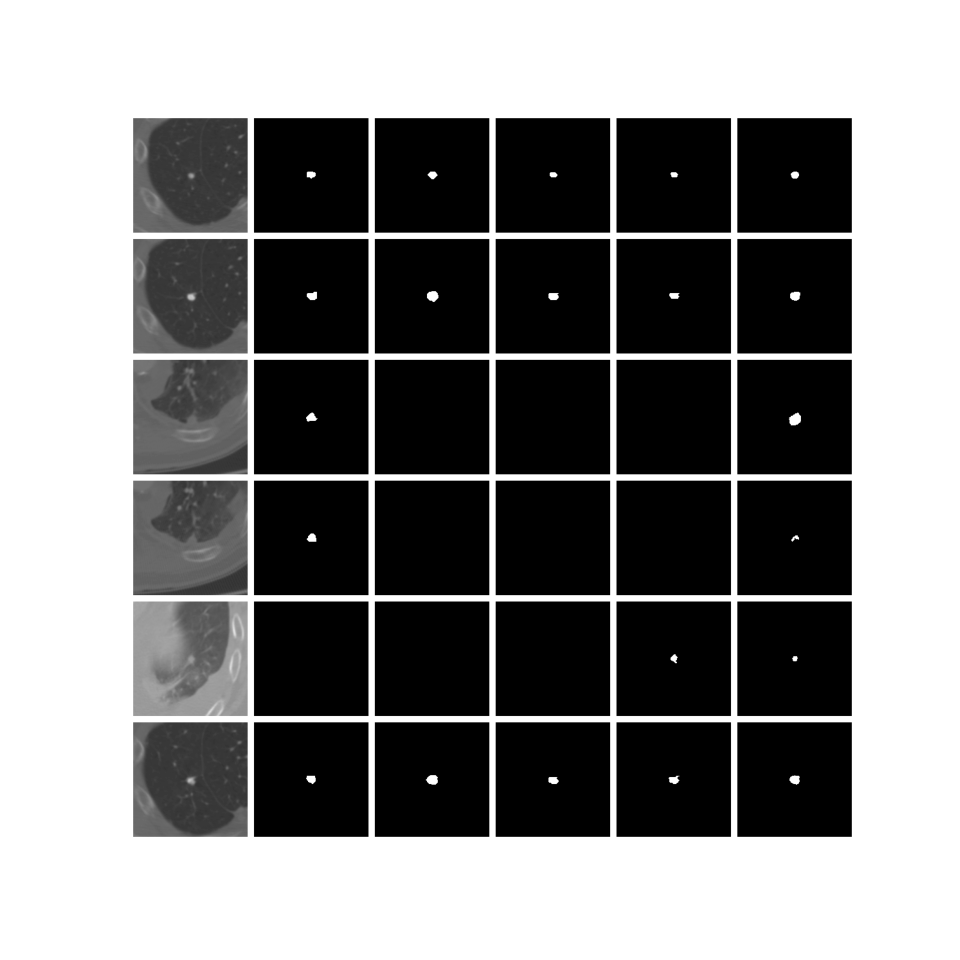}
    \caption{Visualization of 5 Predicted Annotations using Prior Sampling on LIDC Dataset Case 0021.}
    \label{fig:lidc-sample}
\end{figure}  

\section{Network Hyper-parameters}

The hyper-parameters that are being used for training is shown in Table \ref{table:hyperparameters}. During training, we used a combination of cross-entropy loss and the Dice-Similarity coefficient (Dice) loss. The former term can quickly guide the network to learn general segmentation and the latter one comes into effect when learning the fine details. We also used a decayed learning rate scheduler with $1\times 10^{-4}$ decay rate. To balance the magnitude of variance, we used a relatively small $\beta=0.1$ and use the average of KL-divergence from all levels in (\ref{eq:hkl}) to keep the scale consistent with the reconstruction loss.

\begin{table}[h]
\center
\begin{tabular}{|l|r|}
\hline
Parameter & Value      \\
\hline
Batch Size & 24 \\
Learning Rate & 0.05 (Synapse)/ 0.01 (LIDC) \\
Input Size & [224, 224] \\ 
Output Size & [512, 512] \\ 
Momentum & 0.9 \\
Weight Decay & 0.0001 \\
Reconstruction Loss & Cross-Entropy (0.4) + Dice (0.6) \\
Epoch & 150 \\
KL-divergence $\beta$ & 0.1 \\
\hline
\end{tabular}
\vspace{1em}
\caption{Model Training Hyper-parameters.}
\label{table:hyperparameters}
\end{table}

\section{Computational Complexity}

Apart from the performance perspectives, we also want to investigate the performance of the given model. Because of the use of windowed transformer modules, our method is significantly faster than raw transformer-based or convolutional-based models. Compared to the previous state-of-the-art Swin-Unet model, our model\footnote{Evaluation performed on GTX Titan X GPU and Intel Xeon E5-1630 v3 CPU} can achieve higher accuracy with minimal additional cost on the total number of parameters and FLOPs. With the extra cost of 0.02\% (3.6K) parameters, we can improve the overall Dice performance by 0.32\% and overall HD score by 15.64\%, which demonstrated the importance of uncertainty modeling in medical image segmentation tasks.

The complexity of popular models in medical image segmentation is shown in Table \ref{table:complexity}. From the table, we can see that our method holds a small number of parameters as well as a small number of flops. The computational efficiency of the model is largely attributed to the tiny design of the Swin-Transformer block we adopted. With the shifted and patched windows attention module, it is significantly more efficient than its competitors. Although our model is similar to nnUnet in terms of the number of parameters, we have approximately $1\%$ of its flops which greatly accelerate the inference time.

\begin{table}[H]

\center
\begin{tabular}{l|c|c|c}
\hline
Methods                             & \# of Params (M) $\downarrow$         & FLOPs (G) $\downarrow$  & Inference Time (s) $\downarrow$     \\
\hline
TransUnet & $96.07$ & $48.34$ & $26.97$ \\
nnUnet
 & $19.07$ & $412.65$ & $10.28$ \\
CoTr
 & $46.51$ & $399.21$ & $19.21$ \\
ASPP
 & $47.92$ & $44.87$ & $25.47$ \\
SETR
 & $86.03$ & $43.49$ & $24.86$ \\
\hline
Our Method & $20.08$ & $4.92$ & $0.25$\\ 
\hline
\end{tabular}
\vspace{1em}
\caption{Computational Complexity of Synapse Segmentation Models.}
\label{table:complexity}
\end{table}

The small number of flops greatly benefit the application of this model with rapid inference and testing. Using the Swin-T backbone, we can sample in the inference stage while having a reasonable waiting time for patients. Compared to previous methods, we can generate nearly 50 samples before previous models can generate one single prediction. 

\section{Figures for Out-of-distribution Detection}

\begin{figure}[H]
    \centering
    \includegraphics[width=10cm]{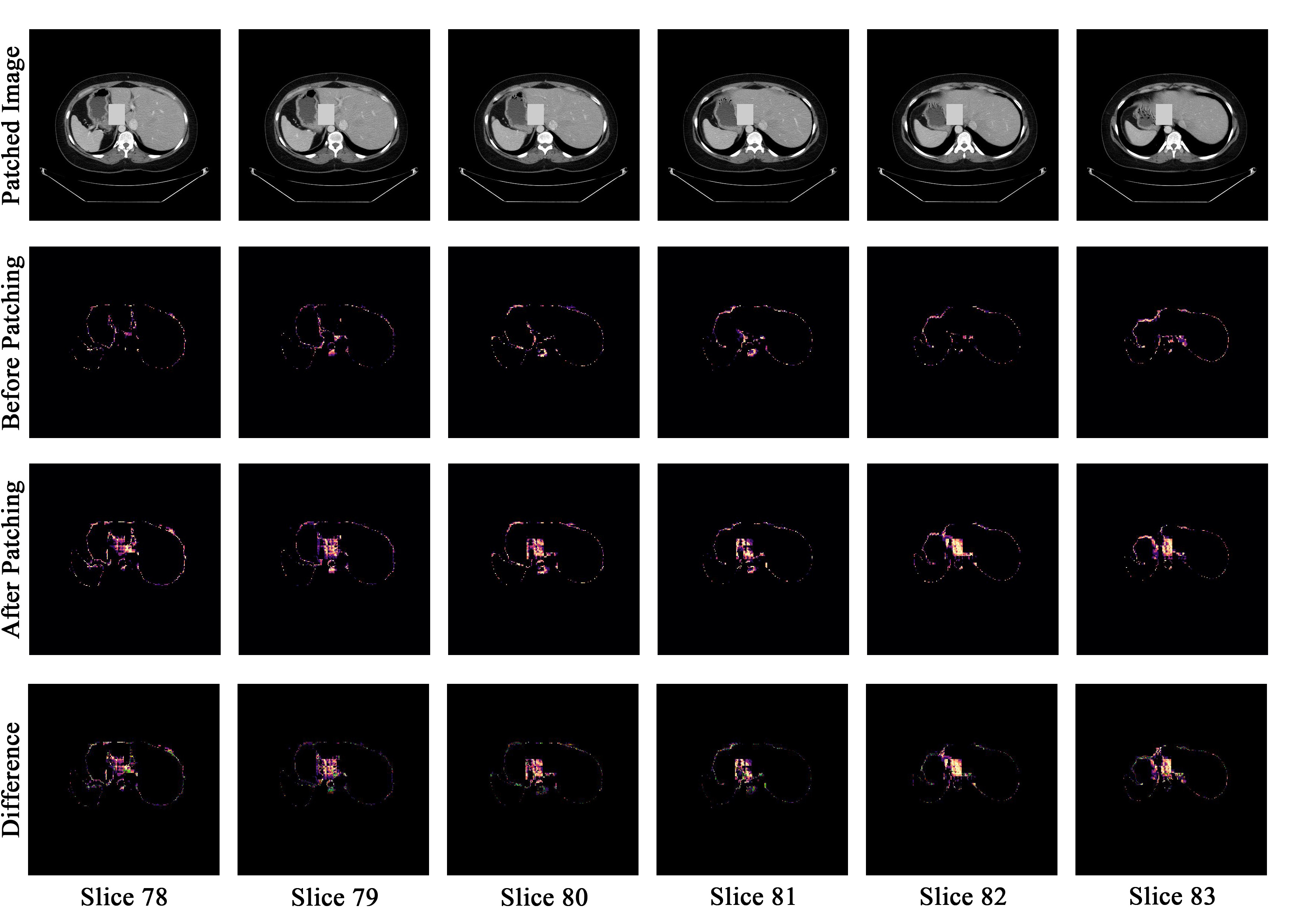}
    \caption{Qualitative Study of Randomly Patched Samples on Synapse dataset Test Case 35. The variance of the predicted 20 samples is visualized in magma colormap.}
    \label{fig:random-patch}
\end{figure}

\begin{figure}[H]
    \centering
    \includegraphics[width=8cm]{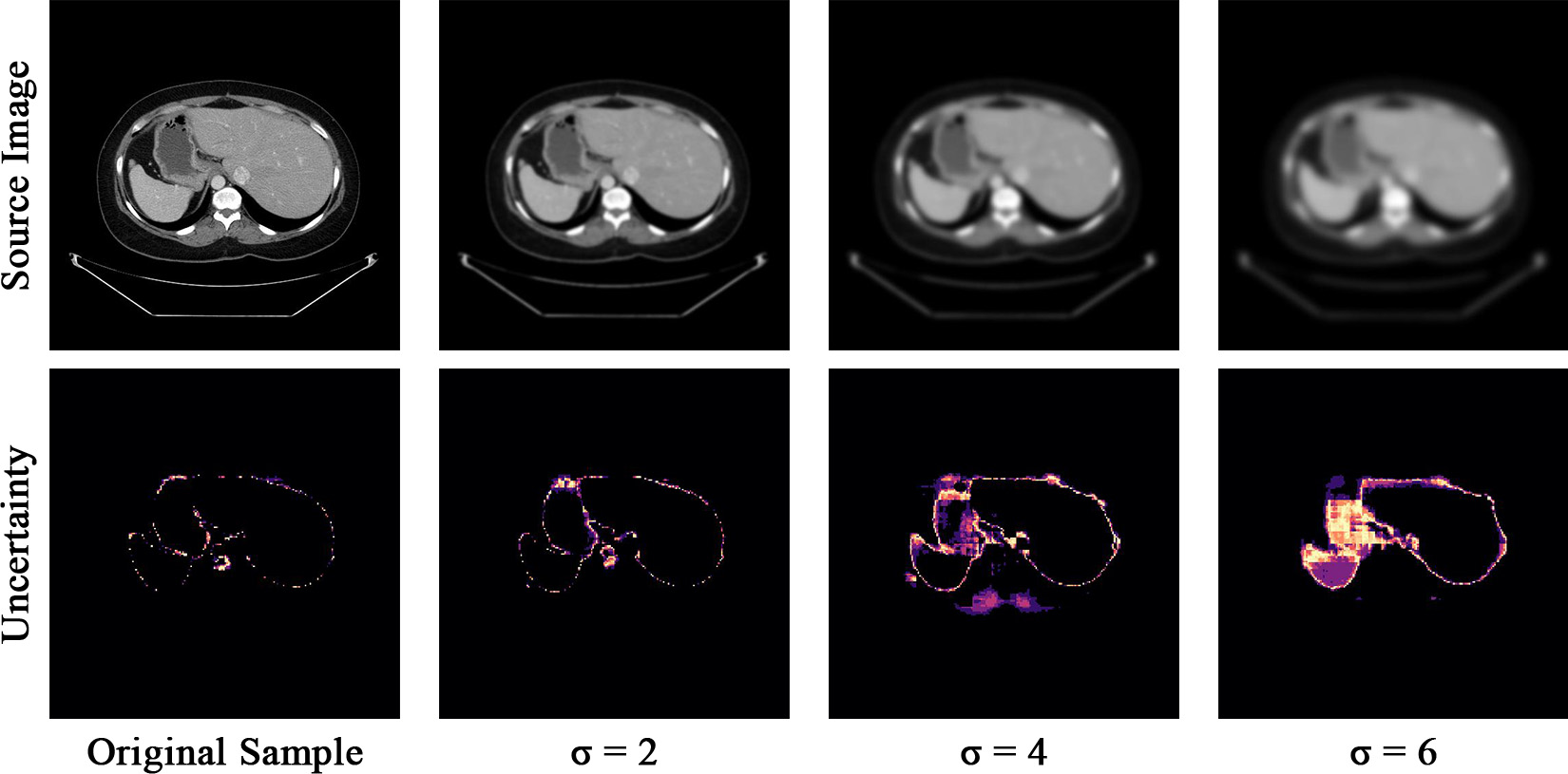}
    \caption{Qualitative Study of the Impact of Gaussian Blur Filters ($\sigma$) on Synapse dataset Test Case 35 (Slice 80). The variance of the predicted 20 samples is visualized in magma colormap.}
    \label{fig:filter}
\end{figure}

\begin{figure}[H]
    \centering
    \includegraphics[width=8cm]{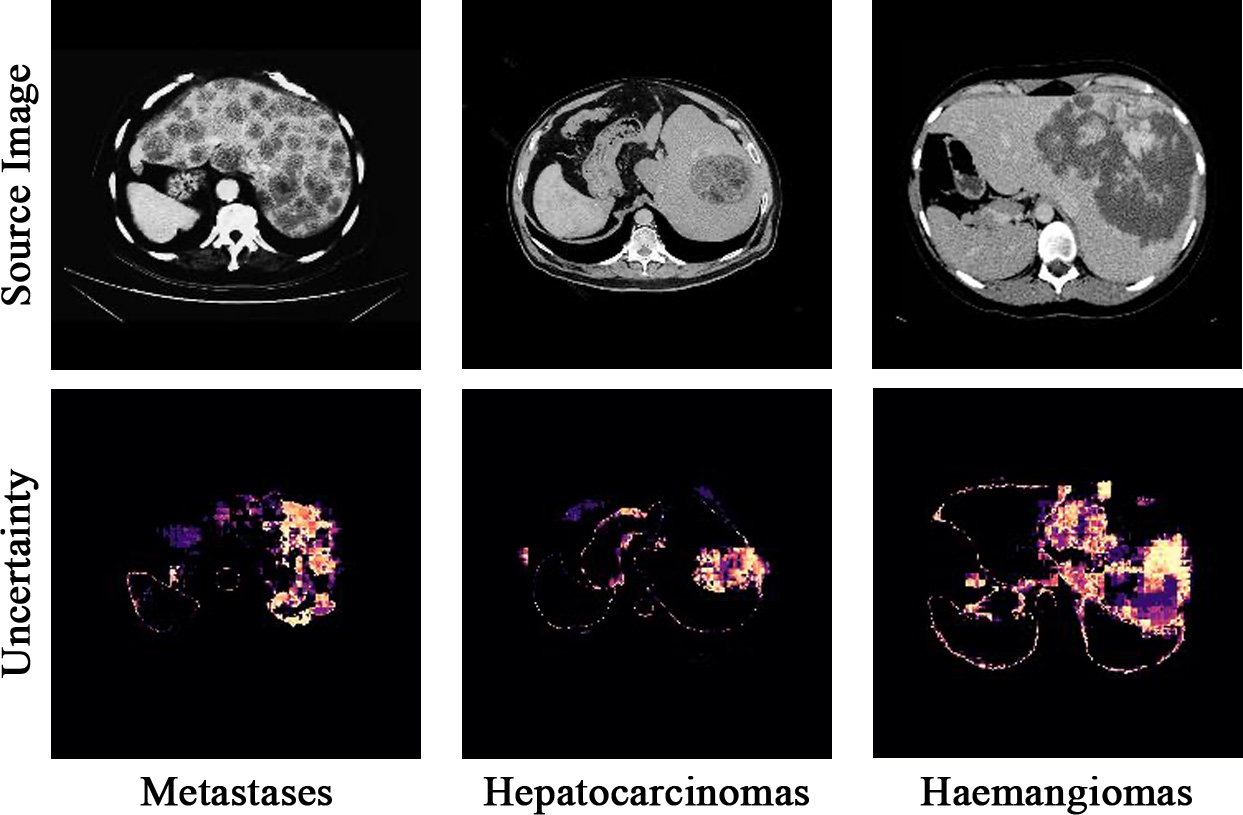}
    \caption{Qualitative Study of Real Liver Tumor. The variance of the predicted 20 samples is visualized in magma colormap.}
    \label{fig:liver-tumor}
\end{figure}

\section{Latent Space Visualization}

\begin{figure}[H]
    \centering
    \includegraphics[width=10cm]{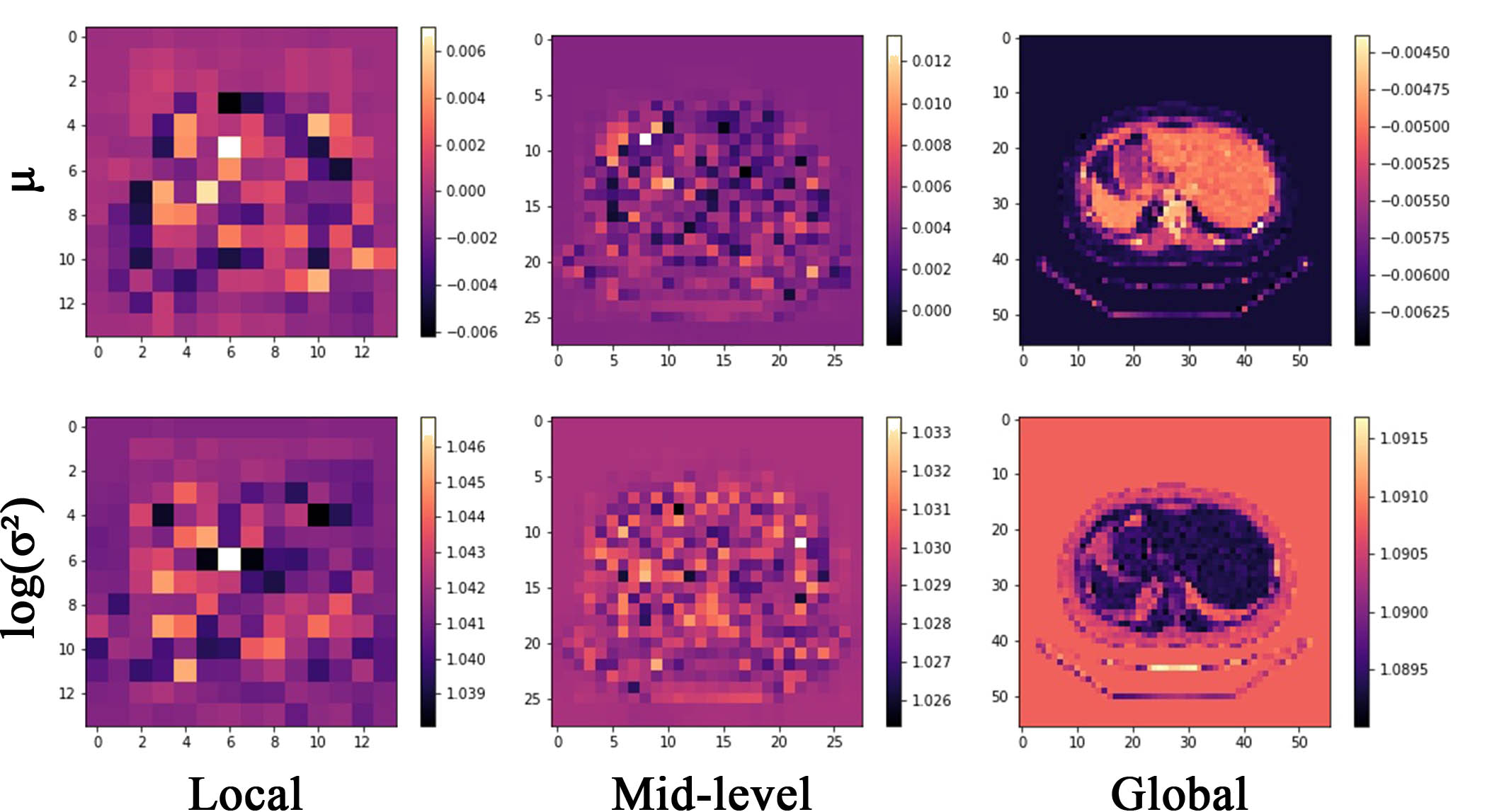}
    \caption{Visualization of mean $\mu$ and  variance $\log(\sigma^2)$ in Gaussian Latent Space across 3 Levels on Synapse dataset
Test Case 35 (Slice 80).}
    \label{fig:uncertainty-level}
\end{figure}

\end{document}